\documentclass[nofootinbib,aps,preprintnumbers]{revtex4-1}

\pdfoutput=1
\usepackage{multirow}
\usepackage{amsmath,amssymb}
\usepackage{graphicx}
\usepackage{color}
\usepackage{url}
\usepackage{hyperref}
\usepackage{natbib}

\begin{document}
\def\contentsname{{\normalsize Content}}
\def\tablename{Table}
\def\figurename{Figure}

\def\pveto{P_\text{veto}}
\def\nj{n_\text{jets}}
\def\meff{m_\text{eff}}
\def\ptmin{p_T^\text{min}}
\def\gtot{\Gamma_\text{tot}}
\def\as{\alpha_s}
\def\az{\alpha_0}
\def\gz{g_0}
\def\w{\vec{w}}
\def\sdag{\Sigma^{\dag}}
\def\s{\Sigma}
\newcommand{\psib}{\overline{\psi}}
\newcommand{\Psib}{\overline{\Psi}}
\newcommand\one{\leavevmode\hbox{\small1\normalsize\kern-.33em1}}
\newcommand{\Mpl}{M_\mathrm{Pl}}
\newcommand{\p}{\partial}
\newcommand{\mat}{\mathcal{M}}
\newcommand{\lag}{\mathcal{L}}
\newcommand{\ord}{\mathcal{O}}
\newcommand{\ope}{\mathcal{O}}
\newcommand{\qqquad}{\qquad \qquad}
\newcommand{\qqqquad}{\qquad \qquad \qquad}

\newcommand{\qb}{\bar{q}}
\newcommand{\matx}{|\mathcal{M}|^2}
\newcommand{\really}{\stackrel{!}{=}}
\newcommand{\msbar}{\overline{\text{MS}}}
\newcommand{\qns}{f_q^\text{NS}}
\newcommand{\lqcd}{\Lambda_\text{QCD}}
\newcommand{\met}{\slashchar{p}_T}
\newcommand{\pmiss}{\slashchar{\vec{p}}_T}

\newcommand{\sq}{\tilde{q}}
\newcommand{\go}{\tilde{g}}
\newcommand{\st}[1]{\tilde{t}_{#1}}
\newcommand{\stb}[1]{\tilde{t}_{#1}^*}
\newcommand{\nz}[1]{\tilde{\chi}_{#1}^0}
\newcommand{\cp}[1]{\tilde{\chi}_{#1}^+}
\newcommand{\cm}[1]{\tilde{\chi}_{#1}^-}
\newcommand{\CP}{CP}

\providecommand{\mg}{m_{\tilde{g}}}
\providecommand{\mst}[1]{m_{\tilde{t}_{#1}}}
\newcommand{\msn}[1]{m_{\tilde{\nu}_{#1}}}
\newcommand{\mch}[1]{m_{\tilde{\chi}^+_{#1}}}
\newcommand{\mne}[1]{m_{\tilde{\chi}^0_{#1}}}
\newcommand{\msb}[1]{m_{\tilde{b}_{#1}}}
\newcommand{\vsm}{\ensuremath{v_{\rm SM}}}

\newcommand{\mev}{{\ensuremath\rm MeV}}
\newcommand{\gev}{{\ensuremath\rm GeV}}
\newcommand{\tev}{{\ensuremath\rm TeV}}
\newcommand{\fb}{{\ensuremath\rm fb}}
\newcommand{\ab}{{\ensuremath\rm ab}}
\newcommand{\pb}{{\ensuremath\rm pb}}
\newcommand{\br}{{\ensuremath\rm BR}}
\newcommand{\sign}{{\ensuremath\rm sign}}
\newcommand{\iab}{{\ensuremath\rm ab^{-1}}}
\newcommand{\ifb}{{\ensuremath\rm fb^{-1}}}
\newcommand{\ipb}{{\ensuremath\rm pb^{-1}}}

\def\slashchar#1{\setbox0=\hbox{$#1$}           
   \dimen0=\wd0                                 
   \setbox1=\hbox{/} \dimen1=\wd1               
   \ifdim\dimen0>\dimen1                        
      \rlap{\hbox to \dimen0{\hfil/\hfil}}      
      #1                                        
   \else                                        
      \rlap{\hbox to \dimen1{\hfil$#1$\hfil}}   
      /                                         
   \fi}
\newcommand{\dslash}{\slashchar{\partial}}
\newcommand{\Dslash}{\slashchar{D}}

\newcommand{\eg}{\textsl{e.g.}\;}
\newcommand{\ie}{\textsl{i.e.}\;}
\newcommand{\etal}{\textsl{et al}\;}

\setlength{\floatsep}{0pt}
\setcounter{topnumber}{1}
\setcounter{bottomnumber}{1}
\setcounter{totalnumber}{1}
\renewcommand{\topfraction}{1.0}
\renewcommand{\bottomfraction}{1.0}
\renewcommand{\textfraction}{0.0}
\renewcommand{\thefootnote}{\fnsymbol{footnote}}

\newcommand{\rig}{\rightarrow}
\newcommand{\lrig}{\longrightarrow}
\renewcommand{\d}{{\mathrm{d}}}
\newcommand{\be}{\begin{eqnarray*}}
\newcommand{\ee}{\end{eqnarray*}}
\newcommand{\gl}[1]{(\ref{#1})}
\newcommand{\ta}[2]{ \frac{ {\mathrm{d}} #1 } {{\mathrm{d}} #2}}
\newcommand{\bee}{\begin{eqnarray}}
\newcommand{\eee}{\end{eqnarray}}
\newcommand{\beeq}{\begin{equation}}
\newcommand{\eeeq}{\end{equation}}
\newcommand{\mc}{\mathcal}
\newcommand{\mr}{\mathrm}
\newcommand{\ep}{\varepsilon}
\newcommand{\emt}{$\times 10^{-3}$}
\newcommand{\emfo}{$\times 10^{-4}$}
\newcommand{\emfi}{$\times 10^{-5}$}

\newcommand{\revision}[1]{{\bf{}#1}}

\newcommand{\hzero}{h^0}
\newcommand{\Hzero}{H^0}
\newcommand{\Azero}{A^0}
\newcommand{\PHiggs}{H}
\newcommand{\PW}{W}
\newcommand{\PZ}{Z}

\newcommand{\sw}{\ensuremath{s_w}}
\newcommand{\cw}{\ensuremath{c_w}}
\newcommand{\swd}{\ensuremath{s^2_w}}
\newcommand{\cwd}{\ensuremath{c^2_w}}

\newcommand{\mhhd}{\ensuremath{m^2_{\Hzero}}}
\newcommand{\mhh}{\ensuremath{m_{\Hzero}}}
\newcommand{\mlhd}{\ensuremath{m^2_{\hzero}}}
\newcommand{\Mlh}{\ensuremath{m_{\hzero}}}
\newcommand{\mad}{\ensuremath{m^2_{\Azero}}}
\newcommand{\mhpd}{\ensuremath{m^2_{\PHiggs^{\pm}}}}
\newcommand{\mhp}{\ensuremath{m_{\PHiggs^{\pm}}}}

 \newcommand{\sa}{\ensuremath{\sin\alpha}}
 \newcommand{\ca}{\ensuremath{\cos\alpha}}
 \newcommand{\cad}{\ensuremath{\cos^2\alpha}}
 \newcommand{\sad}{\ensuremath{\sin^2\alpha}}
 \newcommand{\sbd}{\ensuremath{\sin^2\beta}}
 \newcommand{\cbd}{\ensuremath{\cos^2\beta}}
 \newcommand{\cb}{\ensuremath{\cos\beta}}
 \renewcommand{\sb}{\ensuremath{\sin\beta}}
 \newcommand{\tanbd}{\ensuremath{\tan^2\beta}}
 \newcommand{\cotbd}{\ensuremath{\cot^2\beta}}
 \newcommand{\tanb}{\ensuremath{\tan\beta}}
 \newcommand{\tb}{\ensuremath{\tan\beta}}
 \newcommand{\cotb}{\ensuremath{\cot\beta}}

\newcommand{\GeV}{\ensuremath{\rm GeV}}
\newcommand{\MeV}{\ensuremath{\rm MeV}}
\newcommand{\TeV}{\ensuremath{\rm TeV}}

\title{Spying an invisible Higgs}

\preprint{IPPP/14/101}
\preprint{DCPT/14/201}
\preprint{MCnet-14-23}

\author{Catherine Bernaciak$^{1}$, Tilman Plehn$^{1}$, Peter Schichtel$^{1,2}$, and Jamie Tattersall$^{1}$}

\affiliation{$^1$ Institut f\"ur Theoretische Physik, Universit\"at Heidelberg, Germany}
\affiliation{$^2$ Institute for Particle Physics Phenomenology, Department of Physics, Durham University, UK}

\begin{abstract}
We investigate the potential of multivariate techniques to improve the
LHC search for invisible Higgs decays in weak boson fusion. We find
that in the coming runs the LHC will be able to probe an invisible
Higgs width of 28\% within a year and 3.5\% during a high luminosity
run. A significant improvement over these estimates requires an
analysis of QCD radiation patterns down to 10~GeV. Such an analysis
can improve the reach at the high luminosity run to 2\%. Throughout
our analysis we employ a conservative, data driven background determination.
\end{abstract}


\maketitle
\tableofcontents

\newpage

\section{Introduction}
\label{sec:intro}

The discovery of the Higgs
boson~\cite{Higgs:1964ia,*Higgs:1964pj,*Englert:1964et,*Higgs:1966ev}
on 4th July 2012 was a triumph for the LHC physics
program~\cite{Aad:2012tfa,*Chatrchyan:2012ufa}. After the one missing
piece of the Standard Model (SM) was found, attention quickly turned
to measuring the properties of the particle. So far, all measurements
at the LHC appear to be compatible with the Standard Model. Most
notably, this holds for the measurement of the Higgs couplings to all
Standard Model particles~\cite{Klute:2012pu,*Lopez-Val:2013yba}. Given
the multitude of indirect constraints on the Higgs sector and the
sizable error bars this is not completely unexpected.  With an upgrade
in energy and much more data to collect, it is vital that all of the
possible production and decay modes of the Higgs are probed to as high
accuracy as possible.

The question of whether the Higgs boson has an invisible decay width
is particularly important. In the Standard Model, such an invisible
width is negligible compared to the expected LHC reach.  However, many
theories of new physics predict an invisible decay width competitive
with the total Standard Model
width~\cite{Djouadi:2012zc,*Englert:2014uua}. The main motivation for
searching for such decays is that the Higgs sector could be linked to
a solution to dark matter. More precisely, due to the
super-renormalizable nature of the Higgs mass term, any singlet field
can mix with the Higgs and a portal into a hidden sector
opens~\cite{Patt:2006fw}. The Higgs portal opens a wealth of options
for model building ranging from simple dark matter models to more
complicated unified
models~\cite{Patt:2006fw,*Gonderinger:2009jp,*Bertolami:2007wb,*Andreas:2010dz,*Englert:2013gz}. Whatever
guides the exact composition of such a hidden sector, a Higgs portal
would always show itself through an invisible decay
width~\cite{Shrock:1982kd,*Li:1985hy,*Griest:1987qv,*Low:2011kp,*Belanger:2013kya,*Bento:2001yk,*Englert:2011aa}, which sometimes
leads to somewhat misguided speculation~\cite{Englert:2011us}. A
measurement of an invisible Higgs decay would therefore directly lead
to interesting physics beyond the Standard Model and hopefully be
connected to a viable dark matter candidate.

Recently hints of new physics have appeared in final states with
missing energy at the LHC~\cite{Curtin:2012nn, *Curtin:2013gta,
  *Rolbiecki:2013fia, *Curtin:2014zua, *Kim:2014eva} in the $W^+W^-$
cross-section
and supersymmetric electroweak
searches.
All these
anomalies involve missing energy in final states connected to
electroweak symmetry breaking, which means they motivate improvements
to the current invisible Higgs searches.\bigskip

There are different strategies to detect such a deviation at the LHC.
The classic search strategy for invisible Higgs decays is based on
weak-boson-fusion (WBF) Higgs
production~\cite{Eboli:2000ze,DiGirolamo:2002vwa,*Bai:2011wz,*Ghosh:2012ep}. Boosted
Higgs production in association with a $W$ and $Z$ boson will
significantly add to the LHC
reach~\cite{Choudhury:1993hv,*Godbole:2003it,*Davoudiasl:2004aj,*Okawa:2013hda}. A
search for invisible Higgs decays in $t\bar{t}H$
production~\cite{Gunion:1993jf,*Zhou:2014dba} will be a challenge at
the LHC, both statistically and systematically.  Experimental
measurements in $ZH$ production and weak-boson-fusion Higgs production
have recently been performed by CMS~\cite{Chatrchyan:2014tja} and for
$ZH$ production only by ATLAS~\cite{Aad:2014iia}. For a given
underlying model, global fits of the Higgs couplings can probe
invisible decay modes contributing to the total Higgs width in a
model-dependent
fashion~\cite{Lafaye:2009vr,Englert:2011yb,Espinosa:2012im,*Espinosa:2012vu,*Giardino:2012ww,*Desai:2012qy}.

In this study we analyze the WBF production channel and ask the
question where it can be improved over the original
findings~\cite{Eboli:2000ze}.  We use a multivariate analysis
implemented as a boosted decision tree (BDT) to separate signal and
background and compare it to a normal cut based approach. One of the
central questions is how much information is lost when we employ a jet
veto instead of a comprehensive study of the jet activity.  In
addition, we devise a completely data driven background determination
with essentially no Monte-Carlo extrapolation between control and
signal regions.  Specifically, the $Z\to\nu\nu$ background is modeled
from $Z\to\ell\ell$ events and the $W\to\ell\nu$ background is
determined from events with a lepton and a transverse mass consistent
with a $W$-decay. The remaining systematic uncertainty is associated
with lepton reconstruction and identification probabilities and the
$W$ and $Z$ branching ratios.  This means that the BDT can be trained
on data and then applied to the signal region, which allows us to
safely push the multivariate analysis without worrying about the
effect this may have on the associated systematic error of the
analysis.\bigskip

In Section~\ref{sec:variables} we start by explaining the kinematic
variables entering the multivariate analysis, both with two tagging
jets only and including additional QCD jets.  In the following
Section~\ref{sec:backgrounds} we present our background
determination. Based on those two chapters we systematically test the
prospects during the upcoming LHC runs in
Section~\ref{sec:results}. We estimate the effectiveness of a central
jet veto and find that the most promising path for an improvement of
the WBF analysis is to include more information on QCD jet radiation.

\section{Signal kinematics}
\label{sec:variables}

All signal and background samples were produced by
\textsc{Sherpa}~\cite{Gleisberg:2008ta} with additional matrix element
jets matched using the CKKW algorithm~\cite{Catani:2001cc}. The events
are then passed through the \textsc{Delphes} detector
simulation~\cite{deFavereau:2013fsa} with advanced parameterizations of
the ATLAS electron, muon and tau reconstruction and identification
algorithms given by the \textsc{CheckMate}
tune~\cite{Drees:2013wra}. The final state jets are clustered using
the anti-$k_T$ algorithm in
\textsc{FastJet}~\cite{Cacciari:2005hq,*Cacciari:2008gp,*Cacciari:2011ma}
with $R =0.4$.

To effectively separate signal and backgrounds we use a Boosted
Decision Tree (BDT) available through
\textsc{Tmva}~\cite{Hocker:2007ht} in the \textsc{Root} analysis
framework~\cite{Antcheva:2011zz}. We use a BDT with 400 trees which
each contain 3 layers and extensively test with different Monte-Carlo
parameters to check that the trees are stable and not over-trained.
To calculate the reach of the LHC we use the CLs
prescription~\cite{Read:2002hq} and display our results in terms of
the invisible Higgs width that can be excluded at the 95\% confidence
level.\bigskip

All signal events are first required to pass a common set of trigger
and selection cuts, including at least two tagging jets and missing
transverse energy with
\begin{alignat}{5}
 p_{T,j}  &> 20~(10)~\gev &\qqqquad |\eta_j| &<4.5  \notag \\
 \met &>100~\gev       &\qqqquad \Delta \phi_{\met,j} &> 0.4 \; .
\label{eq:acc_jet}
\end{alignat}
The angular separation of the jets and the missing energy vector helps
to reduce fake missing energy from mis-measured jet momenta. For the
signal selection we veto leptons, as described in the following
Sec.~\ref{sec:backgrounds}. Events including leptons from $W$ or $Z$
decays will correspondingly serve as control regions. We note that while the
two tagging jets may appear too soft to be used as a trigger, we
still require substantial missing energy, $\met >100~\gev$. Consequently, one
or both jets will by definition have significant $p_T$ that can be triggered on. 

In addition to conservatively using jets with $p_T>20~\gev$ we also
present results with $10~\gev$ jets. This allows us to estimate
possible improvements from the jet kinematics or a central jet
veto~\cite{Kleiss:1987cj,*Baur:1990xe,*Barger:1991ib,*Rainwater:1996ud,*Cox:2010ug,*Gerwick:2011tm}.
However, there are significant doubts of how well this technique will
work once pile-up is included.  On the other hand, particle flow has
shown significant promise in managing the effects of pile-up. This can
especially be seen in boosted jet studies.  On the theory side the
number and the kinematic features of $10~\gev$ jets are challenging to
predict and will induce large errors in the analysis. Concerning the
experimental systematics and the theory uncertainty we emphasize that
our backgrounds are determined in a completely data driven way, as
described in Sec.~\ref{sec:backgrounds}. Any source of soft or collinear QCD
radiation will equally affect both the background control regions and
signal regions.\bigskip

The first set of variables we define are the usual variables used in
many WBF analyses along with $\met$ due to the invisible Higgs signal
we are searching for,
\begin{alignat}{5}
\{   p_{T,j}, \; |\eta_{j_1} - \eta_{j_2}|, \; \eta_{j_1}\cdot\eta_{j_2}, \; m_{j_1j_2}, \; \Delta \phi_{j_1,j_2}, \; \met \} 
\qquad \text{(default)} \; . 
\label{eq:zep_var} 
\end{alignat}
From the original analysis of the LHC sensitivity to an invisible
Higgs~\cite{Eboli:2000ze} we quote the cut values with the exception
of a the maximum jet rapidity,
\begin{alignat}{5}
   p_{T,j} &> 40~\gev \qqquad & 
   \met &> 100~\gev \qqquad &
   m_{j_1j_2} &> 1200~\gev \notag \\
   |\eta_j| &< 4.5 \qqquad &
   |\eta_{j_1} - \eta_{j_2}| &> 4.4 \qqquad &
   \eta_{j_1}\cdot\eta_{j_2} &< 0 \; .
\label{eq:zep1}
\end{alignat}
To easily compare the various results we always present the signal to
background ratio after these default cuts. The central jet veto is
defined by vetoing events with a third central jet $\eta_{j_1} >
\eta_{j_3} > \eta_{j_2}$ where
$p_{T,j_3}>20~\gev$~\cite{Kleiss:1987cj,*Baur:1990xe,*Barger:1991ib,*Rainwater:1996ud,*Cox:2010ug,*Gerwick:2011tm}. 
For the box cuts defined above we order the jets in terms of decreasing $p_T$ to 
easily compare and check our results against the original study. 

Finally, the azimuthal angle between the tagging jets with its
peculiar sensitivity to the Lorentz structure of the
event~\cite{Eboli:2000ze,Plehn:2001nj,*Hankele:2006ma,*Hagiwara:2009wt}
is added,
\begin{alignat}{5}
   \Delta \phi_{j_1,j_2} < 1 \, ,  
\label{eq:zep_delphi}
\end{alignat}
and we include these cuts in turn to give three baseline comparison
points.\bigskip

In a systematic analysis of the multi-jet kinematics we can in
principle rely on Fox--Wolfram moments~\cite{Bernaciak:2013dwa,
*Bernaciak:2012nh}. However, the invisible Higgs analysis is
dominated by the 2-jet sample with a central jet veto, which can be
easily described in terms of a few tagging jet observables. We
therefore use a BDT including simply the kinematic variables
\begin{alignat}{5}
 \{ p_{T,j_1}, \; \eta_{j_1}, \; p_{T,j_2}, \; \eta_{j_2}, \; \Delta
 \phi_{j_1,j_2}, \; \met \} 
\qquad \text{(2-jet)} \; .
\label{eq:basis_vec_2}
\end{alignat}
While the actual kinematic variables differ between our 2-jet setup
and the default set in Eq.\eqref{eq:zep_var}, we have checked that in
a multivariate analysis the two sets are equivalent. When looking at
the 3-jet system we add
\begin{alignat}{5}
 \{   p_{T,j_3}, \; \eta_{j_3}, \; \Delta \phi_{j_1,j_3} \} 
\qquad \text{(3-jet)} \; .
\label{eq:basis_vec_3}
\end{alignat}
For this set of variables we use a forward-backward selection where we define $\eta_{j_1} =
\eta_\text{max}$ (the most forward jet) and
$\eta_{j_2}=\eta_\text{min}$ (the most backward jet). Any additional jets in the event
are then ordered in terms of decreasing $p_T$. Consequently, with this selection, the events after a central jet veto 
and the exclusive 2-jet sample are
identical. We have checked that the forward-backward selection of the
tagging jets gives a better reach for the analysis than using a $p_T$ based selection.

At the matrix element level, the above variables without the missing
transverse energy fully describe the system. However, additional soft
radiation will always be present and produce a transverse boost to the
system. This is why we add $\met$ to the set of variables which we
will target with the help of a multivariate analysis setup. In
Sec.\ref{sec:results} we will give a detailed argument why we can
limit ourselves to the 2-jet and 3-jet samples described
above.

\section{Backgrounds} 
\label{sec:backgrounds}

The background to the invisible Higgs signal is dominated by two
sources, $Z\to\nu\nu$ and $W\to\ell\nu$ where the final state lepton
is either outside the detector acceptance or fails to be
reconstructed. The $W$ and $Z$ production process in association with
$n$ jets can be mediated by a pure QCD process radiating the weak
gauge boson, $\sigma \propto \alpha_s^n \alpha$. Alternatively, there
can be an underlying weak process with jet radiation, $\sigma \propto
\alpha_s^{n-2} \alpha^3$. The QCD process will have a significantly
larger rate, while the kinematics of the jet radiation will be more
signal-like for the weak production
process~\cite{Kleiss:1987cj,*Baur:1990xe,*Barger:1991ib,*Rainwater:1996ud,*Cox:2010ug,*Gerwick:2011tm}. Both
QCD multi-jet and $t\bar{t}$ production were found to contribute a
negligible background compared to $W$ and $Z$ production. In
particular the multi-jet background where one jet is highly
mis-measured leading to a large $\met$ signal is very effectively
reduced by the $\Delta \phi_{\met,j}$ cut.

Our signal events have to pass a lepton veto, implemented in
\textsc{CheckMate}~\cite{Drees:2013wra}. The electron reconstruction
uses the ATLAS `loose' working point and is parametrized as a function
of $p_T$ and $\eta$. The tau veto also relies on the ATLAS `loose'
working point as a function of $p_T$ and has roughly 70\% efficiency for
1-prong and 65\% efficiency for 3-prong hadronic tau
decays. Both the $Z\to\nu\nu$ background and $W\to\ell\nu$ background
where the lepton either fails to be reconstructed or falls outside of
the detector acceptance are determined through control regions. To
determine the backgrounds we rely completely on events with a
reconstructed leptonic $Z$ or $W$ decay either after cuts or in a BDT
trained with signal and background events. In both cases the
backgrounds can be understood in detail using real events. We demand
that these control region events pass exactly the same cuts as for the
signal events. Leptons are required to fulfill
\begin{alignat}{5}
 p_{T,e} &> 10~\gev    &\qqqquad |\eta_e|&<2.5 \notag \\
 p_{T,\mu} &> 5~\gev   &\qqqquad |\eta_\mu|&<2.5 \notag \\
 p_{T,\tau} &> 20~\gev &\qqqquad |\eta_\tau|&<2.5 \; .
\label{eq:acc_lep}
\end{alignat}
 For the $W$
background, we require an isolated muon or electron instead of a
lepton veto. In addition, the transverse mass of the lepton and the
missing transverse momentum has to reconstruct the $W$-mass with
$30~\gev < m_T < 100~\gev$. For the $Z$ control region we require a
pair of leptons of same flavor but opposite charge instead of the
lepton veto. Their invariant mass has to reconstruct the $Z$-mass,
$66~\gev < m_{\ell\ell} < 116~\gev$.\bigskip

Previous studies~\cite{Eboli:2000ze,DiGirolamo:2002vwa} reduce the
statistical uncertainty on the backgrounds by kinematically
extrapolating from control regions with large background populations
to signal regions with far smaller backgrounds. Essentially, the
Monte-Carlo prediction is normalized to data in the control region and
it is assumed that the shape is described well enough to extrapolate
to the signal region with far smaller background. However, we find
that the systematic uncertainty associated with such extrapolations
can easily become the dominant error source at the LHC, especially for
the higher luminosity runs. In addition, reliably estimating the size
of this extrapolation uncertainty accurately will be a challenge once
we have to decide if a measurement actually points to invisible Higgs
decays.

Another approach to estimate the $Z\to\nu\nu$ background has been
pioneered by the mono-jet searches at the LHC. It uses a single hard
photon as a template for the $Z$. Again the motivation to include
these events is the larger corresponding cross-section that reduces
the associated statistical error. However, this also comes at the cost
of introducing a non-trivial systematic uncertainty to account for the
different kinematic structures. The limited systematic control can be
studied in the distribution of the number of jets radiated in hard
photon and $Z$ events~\cite{Englert:2011pq}.

We aim to minimize such extrapolations as much as possible
and demand that the control samples with reconstructed leptons
pass the same cuts as for the signal. In fact, the only extrapolation
in our study are the cases when either a neutrino in $Z\to\nu\nu$ or a
lepton in $W\to\ell\nu$ lies outside the detector acceptance. These
events are rare since they are associated with small $\met$, and the
Monte-Carlo extrapolation of the $W$ and $Z$ decays will be very
accurate.\bigskip

The advantage of using a purely data driven background determination
technique with minimal extrapolation is that it offers the perfect
test bed to push multi-variate techniques in the search for new
physics. Since the backgrounds are safely determined from data there
is no danger of over-training on a Monte-Carlo effect that is not
present in reality. We essentially use the background with
reconstructed leptons as a template for the zero-lepton
background. Deviations in the signal data from the background template
give you access to the new physics. The translation between the two
relies on the ratio,
\begin{alignat}{5}
  R = \frac{\br(Z\to\nu\nu)}{\br(Z\to\ell\ell)} \simeq 2.97 \; ,
\end{alignat}
for $\ell = e,\mu$.  Currently, there is a 2\% uncertainty on the
ratio~\cite{Agashe:2014kda} and this is unlikely to improve in the LHC
lifetime.

In addition, there are systematic uncertainties associated with the
lepton identification, reconstruction and acceptance on both the
background and control regions. Here we use the current CMS mono-jet
search~\cite{Khachatryan:2014rra} as guide since the background is
estimated in a very similar way. Currently the systematic uncertainty
ranges around $3\%$ for the $Z\to\nu\nu$ background and around 4\% for the
$W\to\ell\nu$ background. However, this error is dominated by the
statistical error associated with the number of leptons reconstructed
at different energies and angles. We scale this systematic error
by the luminosity for each of the scenarios that we consider.

We would also like to note that the average $\met$ for $Z\to\nu\nu$ background 
events after our BDT has been applied is $\sim 200\GeV$ for all values of the discriminator
used in our analysis. Consequently, we do not require the reconstruction of high energy
leptons that may have larger uncertainties attached.

\section{LHC Reach} 
\label{sec:results}

\begin{table}[t]
\begin{tabular*}{1.0\textwidth}{@{\extracolsep{\fill} }r|cccc|cc} 
\hline
& \multicolumn{4}{c|}{$p_{T,j} > 20~\gev$} & \multicolumn{2}{c}{$p_{T,j} > 10~\gev$} \\ \hline
$\mathcal{L} [\ifb]$   & Eq.\eqref{eq:zep1} & + jet veto  &  + $\Delta\phi_{jj}$ & BDT 2-jets & BDT 2-jets  & + BDT 3-jets \\ \hline
10  	     & 1.02   		   & 0.49			& 0.47	  		& 0.28 	& 0.18  & 0.16 \\
100  	     & 0.49   		   & 0.20			& 0.18	  		& 0.10	& 0.07  & 0.061 \\
3000  	     & 0.25   		   & 0.094			& 0.069	  	        & 0.035	& 0.025 & 0.021 \\
 \hline 
\end{tabular*}
\caption{Exclusion reach in $\br_\text{inv} = \Gamma_\text{inv}/\Gamma_H$ at 95\% CLs
  to an invisible Higgs width at various luminosities and different
  combinations of cuts and multivariate analyses.}
\label{tab:reach}
\end{table}

To compute the prospects of the upcoming LHC runs at $13~\tev$ in
searching for an invisible Higgs produced in weak boson fusion we
start with jets having $p_{T,j}>20~\gev$.  To reproduce the original
analysis of Ref.~\cite{Eboli:2000ze} we apply the box cuts given in
Eq.\eqref{eq:zep1}, add a central jet veto, and finally include the
cut on the azimuthal tagging jet correlation in
Eq.\eqref{eq:zep_delphi}. We show the corresponding LHC reach in the
invisible Higgs branching ratio at 95\% CLs for different integrated
luminosities in the first three columns of
Tab.~\ref{tab:reach}. Unlike in the original paper we do not include
the jet veto survival probability as an external factor, but use our
multi-jet simulations to simulate the jet veto. One of the reasons is
that one of the assumptions in the computation of such survival
properties, namely a Poisson distribution in the number of radiated
jets, is strictly speaking not correct for the signal and for the
electroweak background~\cite{Gerwick:2011tm}. Nevertheless, our signal
and background rates after the box cuts are consistent with the
original work~\cite{Eboli:2000ze}.\bigskip

With $10~\ifb$ at 13~TeV we find that with the classical
weak-boson-fusion cuts we can probe an invisible width at the same
order as the total Standard Model width.  The jet veto is crucial to
the analysis, increasing the reach in the invisible branching ratio by
a factor two.  Adding the $\Delta\phi_{jj}$ cut only marginally improves
the limit at low luminosity.  This is because with only $10~\ifb$ of
data the analysis is statistically limited.  Whilst the
$\Delta\phi_{jj}$ cut improves the signal-to-background ratio, the
significance is hardly affected. Our results for an integrated
luminosity of $10~\ifb$ at $13~\tev$ are similar to the CMS result
(0.49) based on 20~$\ifb$ at 8~TeV, which can be understood by the
rough scaling of the number of events~\cite{Chatrchyan:2014tja}.

As we move to higher luminosity, the reduced statistical error allows
us to probe smaller invisible Higgs widths.  With $100~\ifb$ we expect
to probe an invisible branching ratio around 20\%, while during the
high-luminosity run with $3000~\ifb$ the limit should drop below
7\%. For the high-luminosity option the $\Delta\phi_{jj}$ box cut is
at least as efficient as the central jet veto. This leads us to expect
that in particular this cut will benefit from a multivariate analysis
instead of a statistically limited cut applied to few remaining signal
events.\bigskip

To test the impact of a multi-variate analysis we start with a BDT
that first only contains the 2-jet sample. The corresponding kinematic
variables are listed in Eq.\eqref{eq:basis_vec_2} and include the
missing transverse energy because it is very specifically affected by
soft physics. This approach can be viewed as a multi-variate analysis
after a jet veto. 

\begin{figure}[t]
 \centering \vspace{-0.0cm}
 \includegraphics[width=0.49\textwidth]{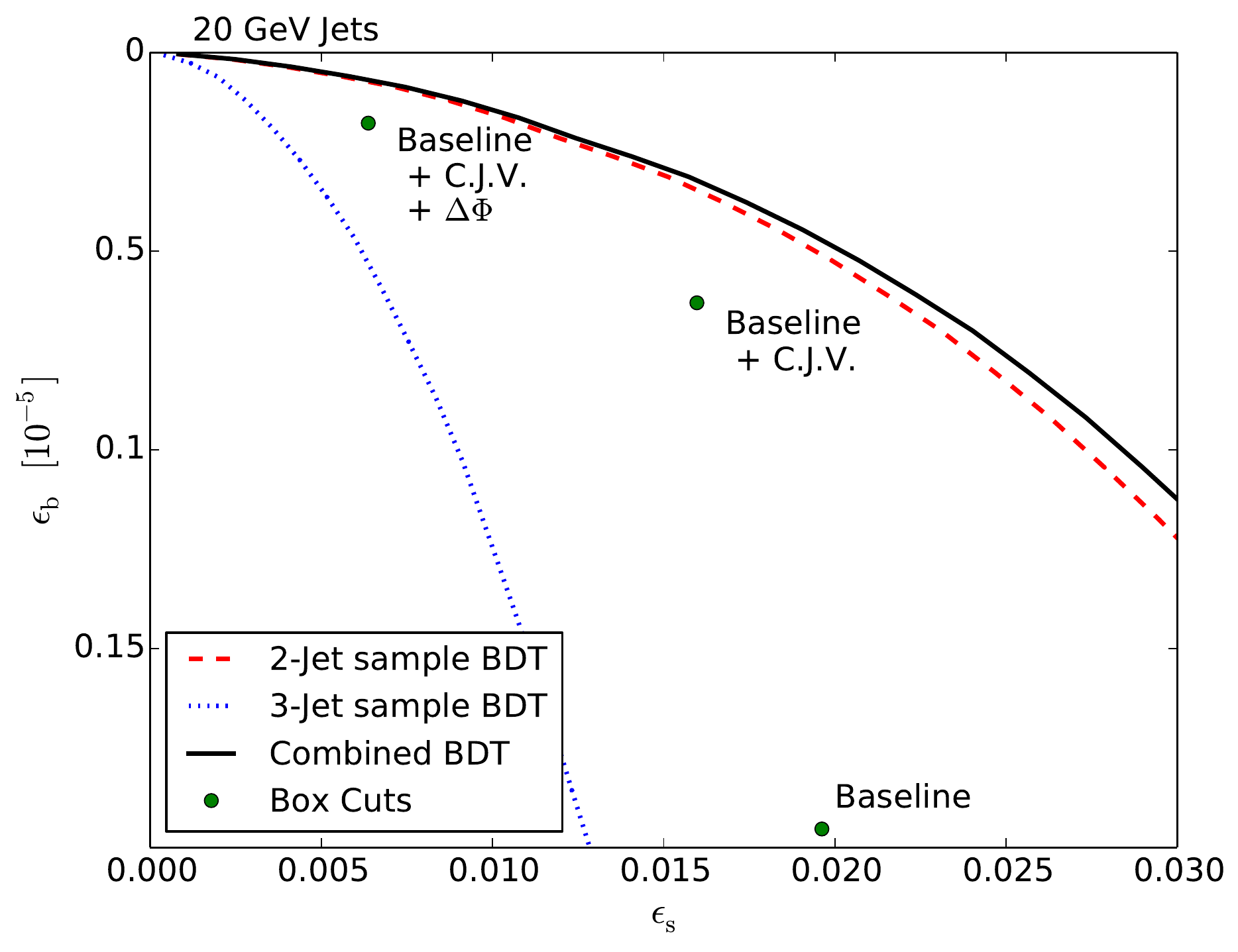}
 \includegraphics[width=0.49\textwidth]{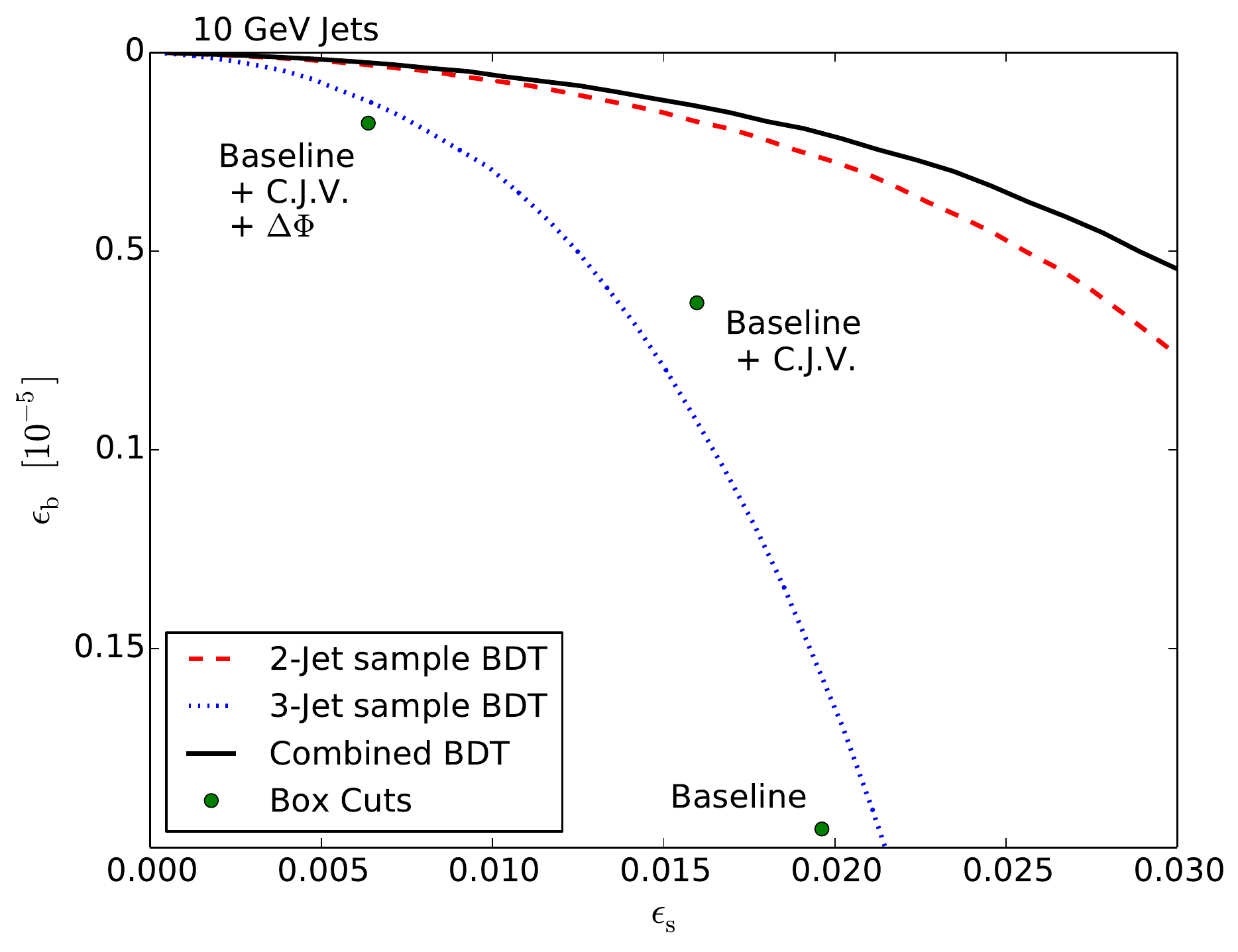}
 \caption{ROC curve for the 2-jet basis vectors,
   Eq.\eqref{eq:basis_vec_2}, and $\met$ on the exclusive $20~\gev$
   (left) and $10~\gev$ (right) 2-jet sample (red dashed), 3-jet, also
   Eq.\eqref{eq:basis_vec_3}, (blue dotted) and the combination
   (black). For comparison the baseline box cut Eq.\eqref{eq:zep1}, 
   also including a jet-veto and in addition a $\Delta
   \phi$ cut, Eq.\eqref{eq:zep_delphi} are shown as points.
 \label{fig:roc}}
\end{figure}

The power of a multi-variate analysis has to be described in terms of
a receiver operator curve (ROC). On this curve we can eventually
choose individual working points.  The ROC curve of our BDT analysis
is optimized to minimize the background for each chosen signal
efficiency separately for the 2-jet and 3-jet samples. In the left
panel of Fig.~\ref{fig:roc} we show the signal efficiency vs the
background fake rate for the 2-jet sample compared to the three box
cut results shown in Tab.~\ref{tab:reach}. For example assuming a
constant signal efficiency we see that the BDT analysis halves the
background fake rate, both compared to the basics cut with the central
jet veto and the basic cuts with the central jet veto and the
$\Delta\phi_{jj}$ cut.

The LHC reach of the multi-variate analysis is then calculated by
including both the statistical and systematic uncertainty on the
background. In fact, the systematic component of the background also
has a statistical component in this analysis since we apply the same
cuts on the background control samples as the background and signal
itself. The ROC point with the most sensitivity for a particular
luminosity is then chosen to calculate the final expected reach. The
result for the 2-jet BDT analysis quoted in Tab.~\ref{tab:reach}
suggests that the reach in the invisible branching ratio improves by a
factor two for all luminosity scenarios.  For the high-luminosity run
an invisible branching ratio down to 3.5\% can be probed.

The remaining key question is to what degree the full information on
additional jets instead of the jet veto improves the LHC reach. For
this purpose we train the BDT analysis on the combined 2-jet and 3-jet
samples, using the extended set of kinematic variables given in
Eq.\eqref{eq:basis_vec_3}.  To see why we only consider the 2-jet and
3-jet sub-samples we examine the signal-to-background ratios in
Tab.~\ref{tab:eff}.  For $20~\gev$ jets, the original $S/B$ values
after the acceptance cuts of Eq.\eqref{eq:acc_jet} decrease with the
number of jets, leading to a statistical limitation of higher jet
multiplicities. Because of the color structure of the signal the
geometric distribution of the jets is very different for the signal
and the background; this difference is most pronounced for the two
tagging jets and gets washed out with any radiated jets. We test this
by comparing the background efficiencies from the BDT for working
point with a constant signal efficiency of $\epsilon_S = 0.01$.  This
point is close to the optimal choice for an integrated luminosity of
$10~\ifb$.  In the corresponding $S/B$ values we indeed observe a
dramatic drop for the 3-jet and 4-jet samples, compared to the same
ratio after acceptance cuts only. If, as it will turn out, the 3-jet
samples do not have significant impact on the final signal extraction
we can safely neglect higher jet multiplicities.

\begin{table}[t]
\begin{tabular*}{1.0\textwidth}{@{\extracolsep{\fill} }l|ccc|ccc} 
\hline
					& \multicolumn{3}{c|}{$p_{T,j} > 20~\gev$} 	& \multicolumn{3}{c}{$p_{T,j} > 10~\gev$} \\ \hline
					& 2-jets 	& 3-jets  	& 4-jets	& 2-jets 	& 3-jets  	& 4-jets \\ \hline
$S/B$ after Eq.\eqref{eq:acc_jet} 	& 1/240 	& 1/360 	& 1/475 	& 1/213		& 1/303		& 1/429   \\
$\epsilon_S$     		  	& 0.01   	& 0.01		& 0.01  	& 0.01		& 0.01  	& 0.01 \\
$\epsilon_B$     			& $1.7 \times 10^{-6}$  	& $1.3 \times 10^{-5}$	& $2.7 \times 10^{-5}$  	& $7.5 \times 10^{-7}$	& $3.2 \times 10^{-6}$  	& $2.4 \times 10^{-5}$ \\
$S/B$     				& 1/2.6   	& 1/21		& 1/42	  	& 1/1.2		& 1/5	 	& 1/38 \\
\hline
\end{tabular*}
\caption{Signal and background efficiencies and signal-to-background
  ratios for different jet multiplicities after the acceptance cuts
  and after applying an optimized BDT discriminant. We choose the BDT
  discriminant such that $\epsilon_S=0.01$.}
\label{tab:eff}
\end{table}

In the left panel of Fig.~\ref{fig:roc} we see that for jets with
$p_{T,j}>20~\gev$ the performance of the 3-jet analysis is
marginal. This is not entirely unexpected.  we know that only a small
fraction of signal events has an additional jet, while the number of
jets in QCD $Z$+jets production is logarithmically enhanced and
follows a Poisson distribution with a maximum at finite jet
multiplicities~\cite{Gerwick:2011tm}. As a matter of fact, this
observation has been the motivation for a central jet
veto~\cite{Kleiss:1987cj,*Baur:1990xe,*Barger:1991ib,*Rainwater:1996ud,*Cox:2010ug,*Gerwick:2011tm}
and can be reproduced from Tab.~\ref{tab:eff}.  Combining the 2-jet
sample with the 3-jet sample shows hardly any improvement.  We only
start to see a slight difference between the 2-jet sample and the full
sample when we look at higher signal acceptances, which is not the
regime which gives us the best reach for invisible Higgs decays for
any of the three luminosity choices.

Using the same BDT setup as in Tab.~\ref{tab:eff} we can determine the
composition of the different backgrounds to the 2-jet sample. We find
that the QCD $W$+jets background is the largest with $\sigma_B =
98~\fb$ for a signal cross section of $\sigma_S = 80~\fb$. QCD
$Z$+jets is the second-largest background with $50~\fb$, followed by
electroweak $Z$+jets and $W$+jets production with $37~\fb$ and
$27~\fb$. This means that for our working point the QCD production
process are slightly larger, but the electroweak processes are hardly
suppressed.\bigskip

The only obvious path to improve the LHC reach for invisible Higgs
decays is to include more information on jet radiation in the
analysis. First, a lower transverse momentum cut on the tagging jets
will increase the signal statistics. Second, the structure of
additional jet radiation will be more distinctive the more jets we
include in the corresponding analysis. In Tab.~\ref{tab:reach} we
compare the $10~\gev$ and $20~\gev$ jet selection for the 2-jet BDT
analysis with the kinematic observables given in
Eq.\eqref{eq:basis_vec_2}. Indeed, the experimentally challenging
analysis setup including softer jets can increase the LHC reach in the
invisible Higgs branching ratio by some 50\%.

In the right panel of Fig.~\ref{fig:roc} we compare the BDT with the
box cuts using $10~\gev$ jets, as before.  In addition to the expected
improvement in the 2-jet analysis we also observe a comparably dramatic
effect on the 3-jet analysis. While it is still not competitive with
the 2-jet analysis, it leads to a significant improvement at large
signal efficiencies of $\epsilon_S > 2\%$. For the optimal working
point an improvement by almost a factor of 2 compared to the $20~\gev$
dominantly 2-jet analysis is shown in Tab.~\ref{tab:reach}.

In this case, where the 3-jet sample does allow for a significant
improvement of the LHC reach we definitely have to see what additional
jets can contribute. In Tab.~\ref{tab:eff} we again show the
signal-to-background ratios for different jet multiplicities after
acceptance cuts and after a BDT analysis of the kinematic
features. While for the $20~\gev$ case the dramatic loss of power
occurred between the 2-jet and 3-jet samples, the softer jets above
$p_{T,j} = 10~\gev$ move this drop to between the 3-jet and the 4-jet
samples. This shift reflects the fact that for sufficiently low
transverse momenta enough signal events will develop additional jet
activity which is different from the corresponding background
patterns. In this case the relevance of the different backgrounds get
re-adjusted: while the dominant background remains QCD $W$+jets
production with $37~\fb$, it is now followed by electroweak $Z$+jets
production with $23~\fb$ and QCD $Z$+jets production with
$19~\fb$. The electroweak $W$+jets channel adds $14~\fb$ to the
combined backgrounds. Electroweak backgrounds exhibit a QCD structure
very similar to the WBF signal, making them more dangerous the more we
rely on jet patterns for the signal extraction.


\section{Conclusions} 
\label{sec:conclusions}

We have investigated how to improve the reach of the LHC to invisible
Higgs decays in the classic weak-boson-fusion
channel~\cite{Eboli:2000ze}. Based on a multi-variate BDT analysis we
found that we can probe signal rates almost twice as small as with
traditional cuts. In particular, at a 13~TeV LHC with $10~\ifb$, we
find that the reach improves from an invisible branching of 47\% to
28\% at 95\%~C.L. For the high-luminosity LHC we expect a final
reach around 3.5\%, significantly benefiting from the increased
statistics. Making use of large expected event samples in the coming
LHC runs we completely rely on reconstructed $W$- and $Z$-decays for
the background simulation, minimizing systematic and theoretical
uncertainties.
 
The central question in our analysis is to what degree a central jet
veto can be improved by taking into account the full information on
the QCD jet radiation. For jets above $20~\gev$ we find that the 3-jet
configuration hardly contributes to the signal extraction unless we
choose a working point with very large signal
efficiencies. Correspondingly, the QCD $Z$+jets and $W$+jets
backgrounds are slightly more dangerous than their electroweak
counterparts.

The main improvement of invisible Higgs searches at the LHC in this
channel needs to incorporate more information on the QCD activity in
the signal and background events. This could be achieved by reducing
the jet threshold to $10~\gev$. While this is clearly not a
conservative requirement on the detectors and the analysis strategy,
methods to include soft jets are being tested for example in jet
substructure studies. With the additional jets the 3-jet topology
does contribute to the signal extraction and should be included beyond
a central jet veto. The high-luminosity run will then be sensitive to
branching ratios of 2.1\%, also limited by our understanding of the
electroweak backgrounds.

\bigskip
\center{\textbf{Acknowledgments}}

We would like to thank Daniel Schmeier for the implementation of the
CLs calculator used in this study. PS acknowledges support from the
European Union as part of the FP7 Marie Curie Initial Training Network
MCnet ITN (PITN-GA-2012-315877) and the IMPRS for Precision Tests of
Fundamental Symmetries.

\bibliography{invisible,selfcites_tp}
\end{document}